\def   \ni {\noindent}

\def   \ssk {\vskip  5truept}

\def   \bsk {\vskip 15truept}

\def   \newline {\hfil\break}

\def   \sssk {\vskip  1truept}

\documentstyle[psfig]{article}
\begin{document}

\hsize 5truein
\vsize 8truein
\font\abstract=cmr8
\font\keywords=cmr8
\font\caption=cmr8
\font\references=cmr8
\font\text=cmr10
\font\affiliation=cmssi10
\font\author=cmss10
\font\mc=cmss8
\font\title=cmssbx10 scaled\magstep2
\font\alcit=cmti7 scaled\magstephalf
\font\alcin=cmr6 
\font\ita=cmti8
\font\mma=cmr8
\def\ref{\par\noindent\hangindent 15pt}
\null


\title{\ni HIGH-ENERGY OBSERVATIONS OF THE BINARY MILLISECOND PULSAR PSR J0218+4232}

\bsk 
\bsk
\author{\ni L.~Kuiper$^{1}$, W.~Hermsen$^{1}$, F.~Verbunt$^{2}$, T.~Belloni$^{3}$, 
A.~Lyne$^{4}$}

\bsk
\affiliation{1) SRON-Utrecht, Sorbonnelaan 2, 3584 CA Utrecht, NL}
\sssk
\affiliation{2) Astronomical Institute, Utrecht University, 3508 TA Utrecht, NL}
\sssk
\affiliation{3) Astronomical Institute ``Anton Pannekoek'', 1098 SJ Amsterdam, NL}
\sssk
\affiliation{4) University of Manchester, NRAL, Jodrell Bank, Cheshire SK11 9DL, UK}
\bsk
\baselineskip = 12pt

\abstract{ABSTRACT \ni We report the detection of pulsed X-ray emission ($4.9\sigma$) from the binary
millisecond pulsar PSR J0218+4232 in a 100 ks ROSAT HRI observation. 
The pulse profile shows a sharp main pulse and an indication for a second weaker pulse at $\sim$ 0.47 
phase separation. The pulsed fraction is 37$\pm$13\%. PSR J0218+4232 was several times in the field of 
view of the high-energy $\gamma$-ray telescope EGRET and a source positionally consistent with the pulsar 
was detected above 100 MeV. Spatial and timing analyses of EGRET data indicate that the source is probably multiple: 
Between 0.1 GeV and 1 GeV PSR J0218+4232 is the most likely counterpart, while the BL Lac 3C66A is the best 
candidate above 1 GeV. If part of the EGRET signal truly belongs to the pulsar, then this would be the {\em first}
millisecond $\gamma$-ray pulsar.}

\bsk
\baselineskip = 12pt

\bsk
\baselineskip = 12pt



\text{\ni 1. INTRODUCTION
\ssk
PSR J0218+4232 was discovered by Navarro et al. (1995) as a highly luminous 2.3 ms radio-pulsar
in a 2.0 day orbit around a white dwarf companion. Its unusually broad radio-pulse profile and the
indication for radio DC-emission points to a nearly aligned rotator. This is supported by Stairs (1998) 
who finds a magnetic inclination angle of  $\sim 10^{\circ}$. At soft X-rays
a source positionally consistent with the pulsar was detected in a 22 ks ROSAT HRI observation by 
Verbunt et al. (1996), who found also indications for pulsed emission. 
Moreover, these authors noticed that the high-energy (HE) EGRET source 2EG J0220+4228 is positionally 
consistent with the pulsar and shows indications for pulsed emission for energies above 100 MeV. 
Here we report the definite detection of pulsed soft X-ray emission from PSR J0218+4232 in a 100 ks 
ROSAT HRI follow-up observation (see Kuiper et al. 1998) and present the results of a more 
detailed analysis using additional EGRET data. 

\ni
     

\begin{figure}[t]
\vspace{0truecm}
\centerline{
\hbox{
\psfig{file=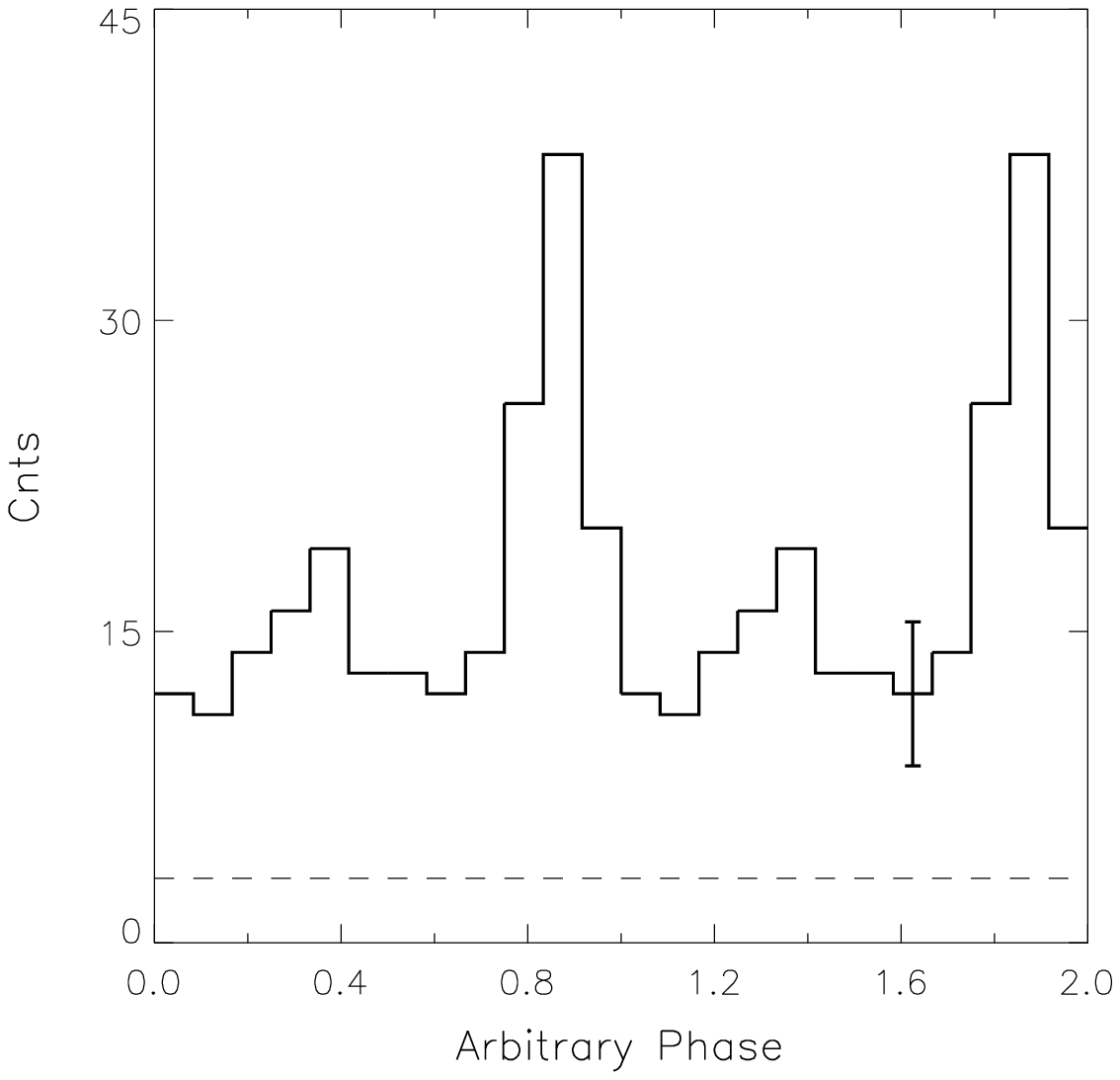,height=5.5cm}
\psfig{file=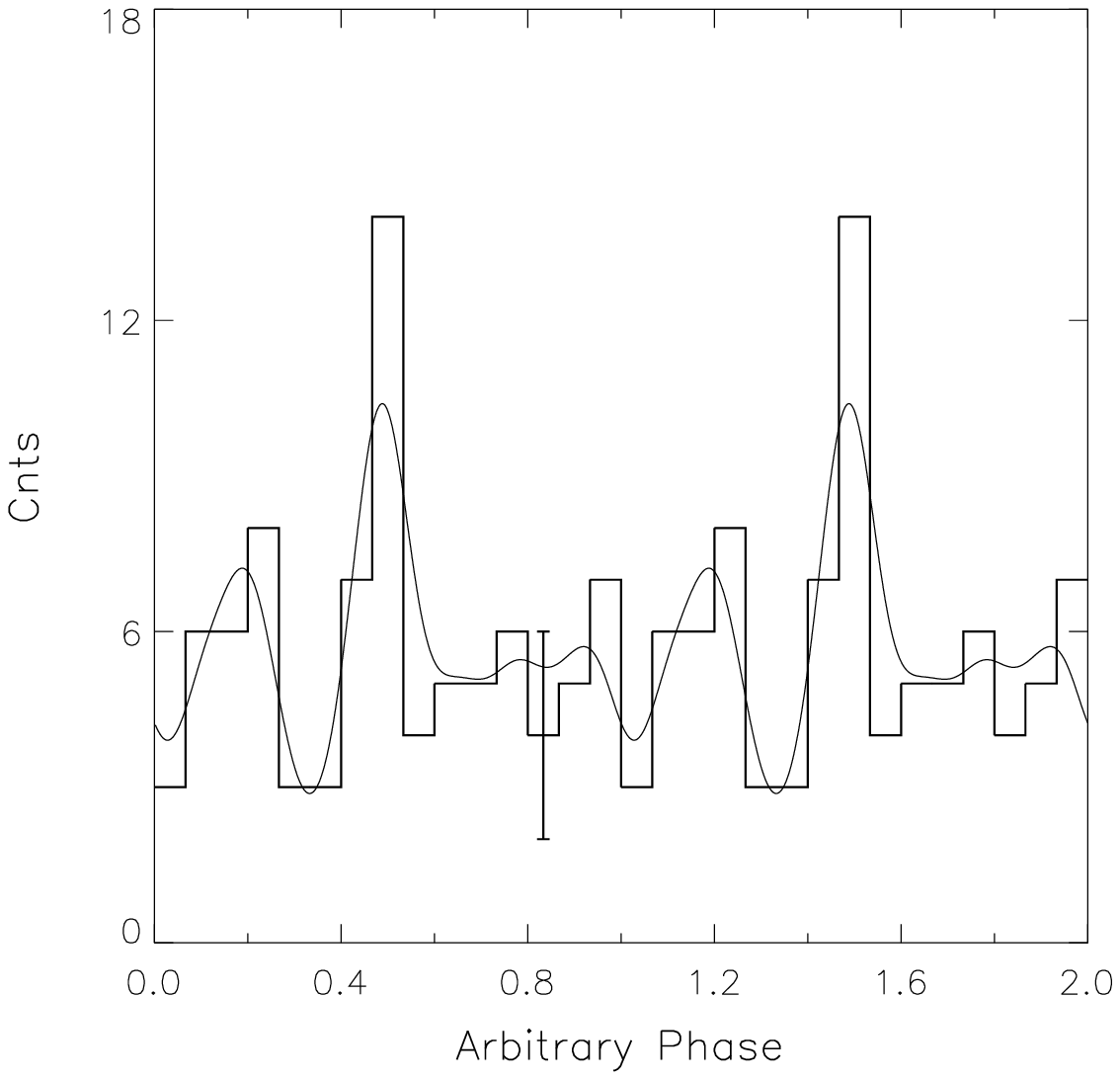,height=5.5cm}
}}
\vspace{0.3truecm}
\caption{FIGURE 1: ({\em left}) Soft X-ray lightcurve of PSR J0218+4232 from a 100 ks 
ROSAT HRI observation applying updated pulsar parameters. The modulation significance is $\sim 4.9\sigma$
($Z_2^2$-test). The spatially determined background level is shown as a straight dashed line. ({\em right}) The 0.3-1 GeV EGRET $\gamma$-ray lightcurve of PSR J0218+4232 combining 
data from VP 15, 211, 325 and 427 with superposed its Kernel Density Estimator (de Jager et al. 1986). 
Extrapolated pulsar parameters (Navarro 1995) are used. The modulation is marginally significant ($2.4\sigma$
in a $Z_4^2$-test). Typical error bars are shown.}
\end{figure}

\bsk
\ni 2. ROSAT HRI OBSERVATION
\ssk
The 100 ks ROSAT HRI observation (0.1-2.4 keV) performed in July 1997 yielded a $21\sigma$ source consistent 
in position with the pulsar. A timing analysis using the pulsar parameters given in Navarro et al. (1995)
extrapolated over a few years yielded a pulsed signal with a $\sim 4.8\sigma$ modulation significance, 
confirming our earlier indications. The pulse profile consists of a sharp main pulse with an indication 
for a second weaker pulse at 0.47 phase separation. The pulsed fraction is $37\pm 13$\% (see Kuiper et al. 1998). 
We produced now a new lightcurve through pulse phase folding with an {\em improved} pulsar ephemeris, determined by one
of us (AL), with a validity interval ranging from MJD 49092 to MJD 50901 ($\sim 5$ years, covering the epoch of our observation). This is shown in Fig. 1 (left) and the morphology is statistically identical to the lightcurve shown 
in Kuiper et al. (1998). 
In Kuiper et al. (1998) we also reported that the large unpulsed component can be explained by emission 
from a compact nebula with a diameter of $\sim 14''$. It is interesting to note that emission from this source 
has also been detected at hard X-rays (2-10 keV) by ASCA in a 40 ks observation (Kawai et al. these proceedings).
This observation time appeared to be too short to detect the weak pulsed signal.

\ni

\begin{figure}[t]
\vspace{0truecm}
\centerline{\hskip 0.35truecm \psfig{file=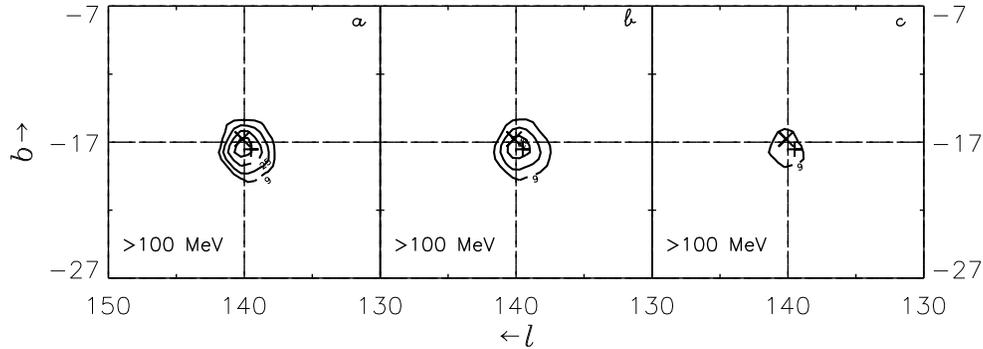,height=4.5truecm}}
\vspace{0.5truecm}
\caption{FIGURE 2. EGRET images for energies above 100 MeV: ({\it a\/}) Total, ({\it b\/}) ``Pulsed'' and
({\it c\/}) ``Unpulsed''. The contours start at a $3\sigma$ detection significance level (1 degree of freedom)
in steps of $1\sigma$. PSR J0218+4232 is indicated by a $+$ sign and 3C66A by a $\times$ sign.}
\end{figure}

\bsk
\ni 3. EGRET OBSERVATIONS
\ssk
PSR J0218+4232 was several times in the field of view of the HE $\gamma$-ray telescope EGRET and a $5.4\sigma$
source, 2EG J0220+4228, positionally consistent with the pulsar was detected above $100$ MeV in the data from
viewing periods (VP) 15 and 211 (Thompson et al. 1995). This source was tentatively identified by e.g.
Dingus et al. (1996) with the BL Lac 3C66A based on its positional coincidence. Verbunt et al. (1996) showed
that indications in the timing analyses make also PSR J0218+4232 a potential counterpart. 
In this paper we have combined EGRET data from more VP's, 15, 211, 325 and 427, and confirmed the source detection 
at a $\sim 6\sigma$ significance level for energies above $100$ MeV.
Timing analyses similar to those presented in Verbunt et al. (1996) have been performed using this enlarged
dataset and again indications for pulsed emission were found at significance levels between $2.5-3\sigma$
for energies above $100$ MeV adopting the extrapolated pulsar parameters presented in Navarro et al. (1995).
The new 0.3-1 GeV lightcurve is shown in Fig. 1 (right). 
The modulation is marginally significant, $2.4\sigma$ in a $Z_4^2$-test. Based on this lightcurve we {\em tentatively} 
defined a ``pulsed'' interval as the combination of the phase ranges $0.08$-$0.24$ and $0.42$-$0.52$.
We also produced images in broad EGRET energy intervals, 0.1-0.3, 0.3-1, 1-10 GeV and the integral band 
$>100$ MeV selecting the events further on their pulse phase.
The maps for energies above 100 MeV in the ``Total'' (no pulse phase selection), ``Pulsed'' and ``Unpulsed''
(complement of ``pulsed'') phase windows are shown in Fig. 2 and the near``ON/OFF'' effect indicates that the 
signal is largely confined to the ``pulsed'' interval suggesting an association with the pulsar.
This ``ON/OFF'' effect is even more pronounced for the differential interval 0.3-1 GeV, but between 1-10 GeV this is 
absent and the most likely counterpart in this energy range is 3C66A (see Fig. 3 (right), confirmed by Hartman et al.
1998). Below 1 GeV PSR J0218+4232 appears to be the most likely candidate both on account of the spatial (see Fig. 3 
left) and timing results. 

\ni

\begin{figure}[t]
\vspace{0.125truecm}
  \centerline{
  \hbox{\hskip 0.5truecm
  \psfig{file=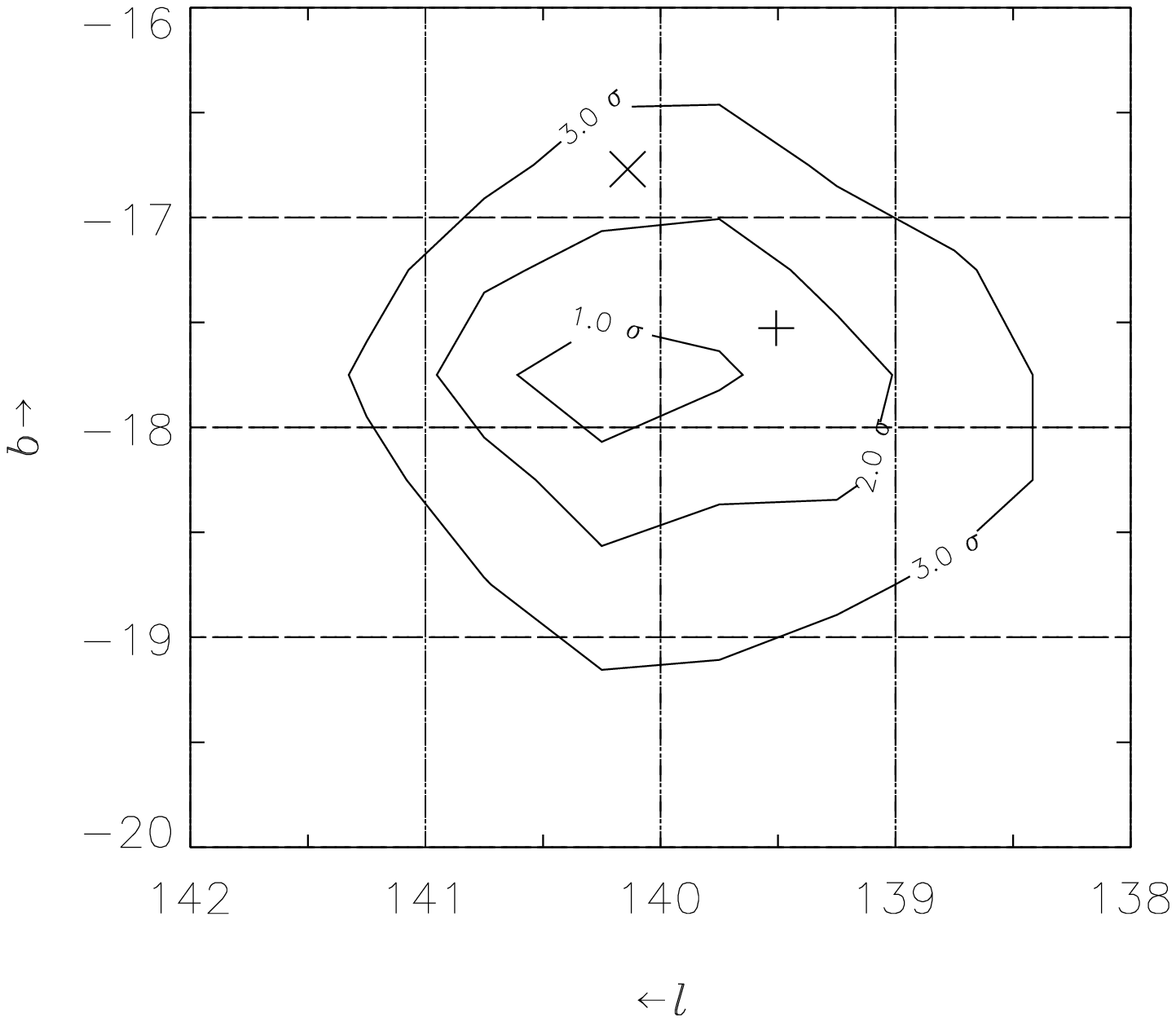,width=5.25cm}
  \hskip -0.25truecm
  \psfig{file=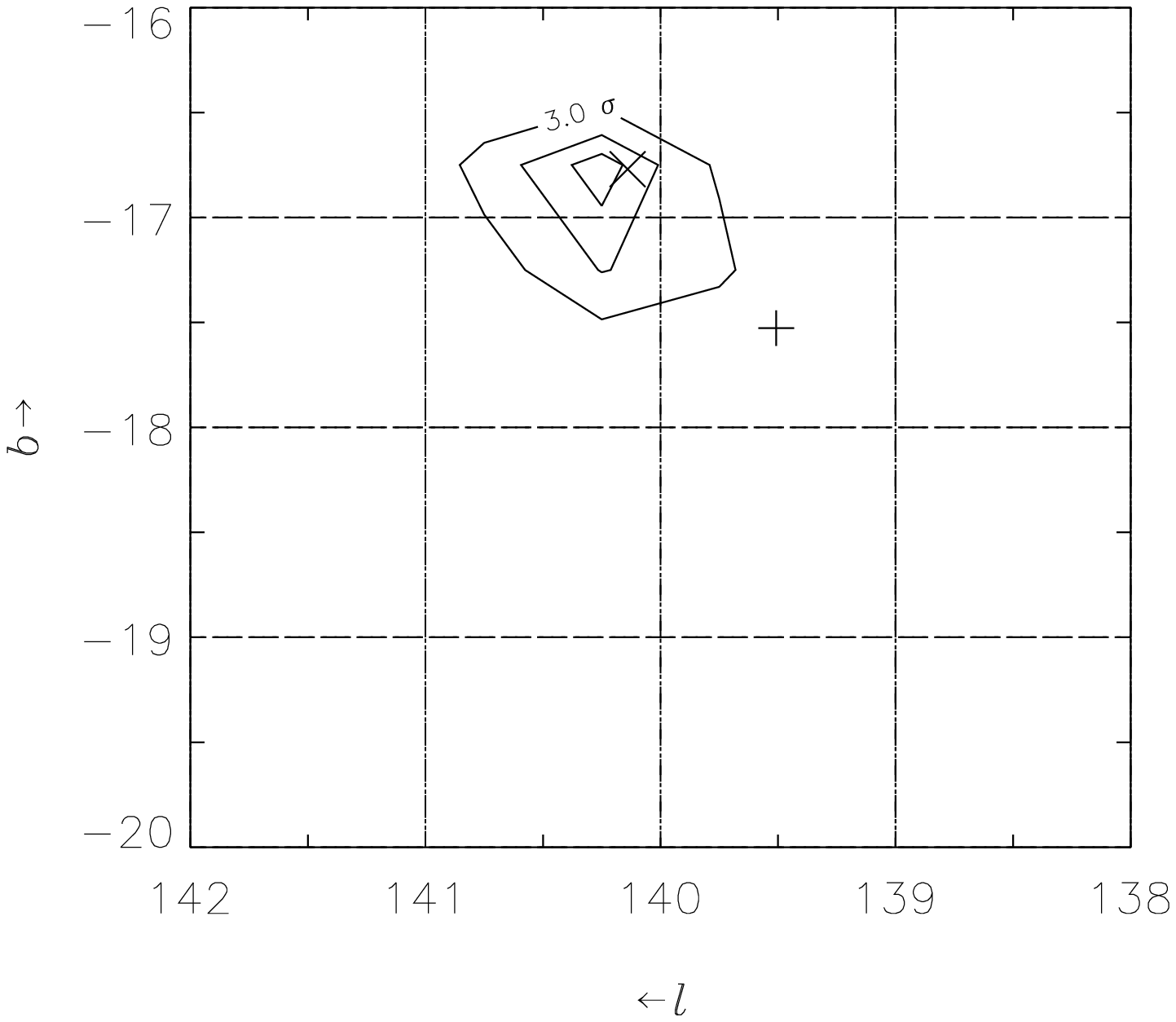,width=5.25cm}
  }
}
\vspace{0.125truecm}

\caption{FIGURE 3 ({\em left}) Location confidence contours for the excess for energies above 100 MeV with the small 
1-10 GeV contribution from 3C66A ``subtracted''. PSR J0218+4232 ($+$) is the most likely candidate although some emission
from 3C66A below 1 GeV can not be excluded. ({\em right}) The same, but now for the 1-10 GeV band. PSR J0218+4232 ($+$) is located outside the $3\sigma$ contour, while 3C66A ($\times$) is consistent with the excess.}

\end{figure}

\bsk
\ni 4. DISCUSSION
\ssk
We have clearly detected pulsed emission from PSR J0218+4232 at soft X-rays (0.1-2.4 keV). The sharp pulses point
to a magnetospheric origin and the most likely production site is near the light cylinder, where the
magnetic field strength $B_{lc}$ is comparable to that of the Crab pulsar. 
The millisecond pulsars PSR B1821-24 and PSR B1937+21 (Kawai et al. these proceedings), both 
ranked in the top 5 in a $B_{lc}$ - ordered scheme, show also sharp X-ray pulses and their spectra appear non-thermal
suggesting also a magnetospheric origin near the light cylinder. 
If the indications for pulsed emission from PSR J0218+4232 
at high energy $\gamma$-rays turn out to be genuine then this would be the {\em first} ms-pulsar detected above $100$ MeV. 
According to Sturner \& Dermer (1994) PSR J0218+4232 should be sufficiently luminous to be detected at HE $\gamma$-rays, if
these are produced in an ``outer gap'' near the light cylinder. Moreover, in a simple spindown-flux ranked scheme, sofar 
very successfull in selecting promising $\gamma$-ray pulsars, PSR J0218+4232 can be found in the top 40, just below the 
established $\gamma$-ray pulsar PSR B1055-52.
A 3-week EGRET observation scheduled for October 1998 might establish the timing signature at HE $\gamma$-rays 
and approved observations with SAX and AXAF will disclose detailed properties of the pulsar and the extended/DC source.  
\ni


\bsk
\baselineskip = 12pt


{\references \ni REFERENCES
\ssk
\ref Dingus, B.L., Bertsch, D.L., Digel, S.W., et al., 1996, ApJ 467, 589
\ref Hartman, R.C., et al., 1998, ApJS in press (3rd EGRET catalogue)
\ref de Jager, O.C., Swanepoel, J.W.H., Raubenheimer, B.C., et al., A\&A 170, 187
\ref Kuiper, L., Hermsen, W., Verbunt, F., et al., 1998, A\&A 336, 545
\ref Navarro, J., de Bruyn, A. G., Frail, D. A., et al., 1995, ApJ 455,L55
\ref Stairs, I. H., 1998, Thesis Princeton University
\ref Sturner, S., Dermer, C., 1994, A\&A  281, 314
\ref Thompson, D.J., Bertsch, D.L., Dingus, B.L., et al., 1995, ApJS 101, 259
\ref Verbunt, F., Kuiper, L., Belloni, T., et al., 1996, A\&A 311,L9
}

\end{document}